\documentclass[twocolumn,           
               showpacs,            
               preprintnumbers,     
               aps,                 
               prd,                 
               letter,             
               superscriptaddress,  
               nofootinbib,         
               tightenlines,        
               floats,floatfix,longbibliography     
               ]{revtex4}

\usepackage{graphicx}
\usepackage{subfigure}
\usepackage{latexsym}
\usepackage{blindtext}
\usepackage{amsmath,amssymb}        
\usepackage[draft=false]{hyperref}
\usepackage{threeparttable}
\usepackage{url}
\usepackage{xcolor}
\usepackage{hyperref}

\newcommand{\PC}[1]{\ensuremath{\left(#1\right)}} 

\input colordvi

\begin{document}



\title{Reconstructing thawing quintessence with multiple datasets}
\author{Nelson A.~Lima}
\affiliation{Institute for Astronomy, University of Edinburgh, Royal Observatory, Blackford Hill, Edinburgh, EH9 3HJ, United Kingdom}
\author{Andrew R.~Liddle}
\affiliation{Institute for Astronomy, University of Edinburgh, Royal Observatory, Blackford Hill, Edinburgh, EH9 3HJ, United Kingdom}
\author{Martin Sahl\'en}
\affiliation{BIPAC, Department of Physics, University of Oxford, Denys Wilkinson Building, 1 Keble Road, Oxford OX1 3RH, UK}
\author{David Parkinson}
\affiliation{School of Mathematics \& Physics, University of Queensland, Brisbane QLD 4072, Australia}
\date{\today}
\pacs{98.80.-k \hfill arXiv:1501.02678}
\preprint{arXiv:1501.02678}


\begin{abstract}
In this work we model the quintessence potential in a Taylor series expansion, up to second order, around the present-day value of the scalar field. The field is evolved in a thawing regime assuming zero initial velocity. We use the latest data from the {\it Planck} satellite, baryonic acoustic oscillations observations from the Sloan Digital Sky Survey, and Supernovae luminosity distance information from Union$2.1$ to constrain our models parameters, and also include perturbation growth data from the WiggleZ, BOSS and the $6$dF surveys. The supernova data provide the strongest individual constraint on the potential parameters.
We show that the growth data performance is competitive with the other datasets in constraining the dark energy parameters we introduce. We also conclude that the combined constraints we obtain for our model parameters, when compared to previous works of nearly a decade ago, have shown only modest improvement, even with new growth of structure data added to previously-existent types of data. \end{abstract}

\maketitle


\section{Introduction}

Dark energy is one of the great discoveries of the 20th century, and remains one of the greatest scientific puzzles of the 21st century. While the observational evidence does not favour any particular physical mechanism beyond a cosmological constant, there are many ideas for mechanisms other than the simplest concordance $\Lambda$CDM model \cite{2014arXiv1401.0046M}. The possibility remains that the observed acceleration is caused by a scalar field, quintessence (see Refs.~\cite{2013CQGra..30u4003T,2013PhRvD..87h3505C} and references therein).
The quintessence field affects both the background expansion history of the Universe, and the rate at which matter over-densities grow. A key area in current and future dark energy (DE) observations is the combination of data on the background expansion history and the structure-growth history \cite{2013arXiv1309.5385H}. The relation between these two histories is sensitive to the properties of the dark energy mechanism, and could allow stronger model constraints than either method separately. 

In this work, assuming quintessence is a valid description of observations, we derive new constraints on the quintessence self-interaction potential and dynamical evolution, based on the latest observational data on baryon acoustic oscillations encoded in the galaxy distribution, the cosmic microwave background, and Type Ia supernovae. We also include galaxy survey data measuring the growth of matter perturbations. 
Earlier work has not included growth-of-structure data (e.g. Refs.~\cite{2005PhRvD..72h3511S, 2007PhRvD..75b3502S,hutererquint1,lamdadevandrew}, though some recent works have placed constraints on inverse power-law quintessence models using new growth-of-structure data \cite{2013arXiv1312.5285P, 2014arXiv1406.0407A}). Our work is distinct in that it uses a direct modelling of the scalar field potential as a Taylor expansion and the field dynamics, rather than relying on indirect parameterizations. We restrict ourselves to the case of a 'thawing' quintessence \cite{2005PhRvL..95n1301C} field with zero initial velocity at early times.

The coming decade will see a series of observational projects, such as the Dark Energy Survey \cite{2010JPhCS.259a2080S}, BOSS \cite{2013AJ....145...10D}, BigBOSS \cite{2012SPIE.8446E..0QM}, {\it WFIRST} \cite{2013arXiv1305.5422S}, {\it Planck} \cite{2006astro.ph..4069T}, HETDEX \cite{2012AAS...21942401H}, DESI \cite{2013arXiv1308.0847L}, {\it Euclid} \cite{2011arXiv1110.3193L} and LSST \cite{2009arXiv0912.0201L}, all of which will probe the growth of structure by means of different techniques. Some of the prospects for constraining quintessence models with these are discussed in Refs.~\cite{2014arXiv1401.0046M,2013arXiv1309.5385H,2014JCAP...03..045T}.

\section{Formalism}

\subsection{Cosmological model}

We follow the approach pioneered in Refs.~\cite{2005PhRvD..72h3511S,2007PhRvD..75b3502S}. We briefly review the set-up here, which is fairly straightforward. We assume that the quintessence field $\phi$ has a self-interaction potential $V(\phi)$, that we expand in a series about the present value of the field which is taken (without loss of generality) to be zero. The quintessence field obeys the equation
\begin{equation}
\ddot{\phi} + 3 H \dot{\phi} = - \frac{dV}{d\phi} \,,
\end{equation}
with the Hubble parameter $H$ given by the Friedmann equation
\begin{equation}
H^2 = \frac{8\pi G}{3} \left(\rho + \rho_{\rm \phi}
\right) \,.
\end{equation}
Here $\rho$ represents all the usual material components: dark matter, baryons, neutrinos, and radiation. We will use $\rho_{{\rm m}}$ to give the matter density (dark matter plus baryons) and $\rho_{\rm \phi} = \dot{\phi}^2/2 + V(\phi)$ the quintessence density. We assume spatial flatness throughout (as motivated by cosmic microwave background (CMB) measurements and the inflationary paradigm), though the generalization to the non-flat case would be straightforward. Our computations start at a redshift of one million.

An important quantity, which determines the cosmological effects we consider from the quintessence field, is the equation of state
\begin{equation}
w_{\rm \phi} \equiv \frac{p_{\rm \phi}}{\rho_{\rm \phi}} =
\frac{\dot{\phi}^2/2 - V(\phi)}{\dot{\phi}^2/2 + V(\phi)} \,.
\end{equation}
In order to have $w_{\rm \phi}$ close to $-1$, mimicking $\Lambda$CDM, the evolution of the field should be potential dominated, which is usually called the slow-roll regime. We focus on thawing quintessence, where the field is initially frozen due to high damping from the Hubble friction term, meaning its energy density is fairly constant at early times. However, at late times, the field starts thawing and slowly rolling on its potential towards larger values of $w_{\rm{\phi}}$ \cite{2013CQGra..30u4003T}.


\subsection{Parameterizations and priors}

\subsubsection{Power series}

We will use a Taylor expansion power series to model the potential $V(\phi)$, as
\begin{equation}{\label{taylorpot}}
V(\phi) = V_0 + V_1\phi + V_2\phi^2 + \ldots,
\end{equation}
and work in units of the reduced Planck Mass, $M_{\rm{P}} = 1/\sqrt{8 \pi G}$, defining $\phi$ in dimensionless units. To have viable flat models, we set $V_{0} = \rho_{\rm{\phi}}^{0}/\rho_{\rm{cri}}^{0} \equiv \Omega_{\rm{\phi}}$, which will be derived from the present-day energy density of matter, $\Omega_{\rm{m}}$, due to the spatial flatness condition. Note that $\Omega_{\rm{m}}$ is the sum of the present-day cold dark matter density and baryon density divided by the present-day critical density, $\Omega_{\rm{c}}$ and $\Omega_{\rm{b}}$ respectively. $\Omega_{\rm{m}}$ also includes the contribution of massive neutrinos, for which we fix $\sum m_{\nu} = 0.06 \hspace{1 mm} \rm{eV}$.

For all our models we have set $\dot{\phi} = 0$ at the initial redshift, $z_i \approx 10^{8}$, so that the scalar field starts at rest. We perform a simple binary search to determine the right $\phi(z_{\rm i}) \equiv \phi_{\rm i}$ that allows us to recover a flat cosmology today. We then rescale the Taylor series parameters of the potential by the value of $\phi (z = 0) \equiv \phi_{\rm f}$ so that the present-day value of $\phi$ is set to zero. As an example, for the first-order series expansion of $V(\phi)$, this means that
\begin{eqnarray}
\label{rescaledparams}
 &\tilde{V}_{0}& = V_{0} + V_1 \phi_{\rm f}, \nonumber \\
 &\tilde{V}_{1}& = V_{1}, \nonumber \\
 &\tilde{\phi}_{\rm i}& = \phi_{\rm i} - \phi_{\rm f} \nonumber 
\end{eqnarray}
and so on for higher-order expansions of the potential. Hence, $\tilde{V}_{0}$ will correspond to the present-day value of the quintessence potential. Effectively, we will be constraining the rescaled parameters, which we indicate with the tilde.


\subsubsection{Priors}

In this work, as detailed before, we expand the quintessence potential in a Taylor series expansion. We limit ourselves to a second-order expansion, which means we will have three dark energy parameters describing our model, apart from the cosmological parameters such as $\omega_{\rm{c}}$ and $\ln \PC{10^{10}A_{\rm{s}}}$. For these parameters we have kept the same prior ranges used in the {\it Planck} analysis \cite{planckres}, which are default for CosmoMC \cite{cosmomc}. We treat CMB foreground parameters identically to Ref.~\cite{planckres}.

In the previous section we defined the zeroth-order term as $V_{0}$. This is constrained by $\Omega_{\rm{m}}$ to guarantee models with a flat cosmology and when only this term is present it corresponds to the present-day energy density of our dark energy component. If the potential is not flat there will be an additional contribution to the present field energy density from its kinetic energy, which data will constrain to be fairly small corresponding to $w_\phi$ close to $-1$.

We have imposed a flat prior on $V_1$ in the range $[-4,4]$ for the first order expansion of the potential. The reason for this is that this range includes, conservatively, the one considered in previous similar works \cite{2005PhRvD..72h3511S, 2007PhRvD..75b3502S}. Given the time elapsed since those works were done, we expect that our range won't have a direct impact on the results and should be quite constrained by the data.

Lastly, there is the second-order term, $V_{2}$. The previous works we refer to were not able to detect significant constraints on this parameter. After some inspection, we believe our results for this parameter are also very prior related, for reasons we will discuss later. Hence, without any particular motivation, we have set a flat prior ranging between $[-10,10]$. For the $V_{1}$ parameter, we reduced the prior to $[-2,2]$, in order to avoid fast rolling of the scalar field for large values of $V_{2}$, which could jeopardize the success of the numerical evaluation of the cosmological evolution predicted by our models.

\section{Data analysis}

\subsection{Observables}

The analysis done in this work was performed using the publicly-available CosmoMC code \cite{cosmomc}. For that, we modified the corresponding quintessence module. For our analysis, we have considered the first {\it Planck} data release \cite{planckres} plus the {\it WMAP} polarization data \cite{wmap9}. We have included CMB lensing in our analysis, as well as the SPT \cite{spt} and ACT \cite{act} data. We have also included the baryonic acoustic oscillations data from the seven-year and nine-year release data sets from the Sloan Digital Sky Survey (SDSS) \cite{sdss7,sdss9}. Furthermore, we use the Union $2.1$ $580$ SNIa catalogue from the Supernova Cosmology Project (SCP) \cite{supernova}, where we conservatively include systematic errors and marginalize over $H_{0}$ as detailed in the appendix of Ref.~\cite{h0marg}. 

Lastly, we include growth of structure data from the WiggleZ Dark Energy Survey \cite{wigglez}, the BOSS survey \cite{bossgrowth} and the $6$dF Galaxy Survey \cite{6dfgrowth}. The two-dimensional power spectrum data from these surveys were used to fit for both the redshift-space distortion effect (which measures the rate of growth of structure $f\sigma_8$) and the Alcock--Paczynski distortion of the survey geometry. The module we use compares our predictions against both types of measurement. Further details of these measurements can be found in Ref.~\cite{wigglezgrowth}.

\subsection{Parameter estimation}

The parameter space we study will be
\begin{equation}
\mathbf{\Theta} =  \left\{100 \, \theta_{\rm{MC}}, \tau, \omega_{\rm{c}}, \omega_{\rm{b}}, \ln \PC{10^{10}A_{\rm{s}}}, n_{\rm{s}}, V_{1}, V_{2} \right\} \,,
\end{equation}
where the cosmological parameters have the same meaning as in Ref.~\cite{planckres}.
The parameter estimation is carried out
using an MCMC approach. The posterior probability of the parameters
$\mathbf{\Theta}$, given the data and a prior probability
distribution $\Pi(\mathbf{\Theta})$, is
\begin{equation}
P(\mathbf{\Theta} | {\rm data}) = \frac{\Pi(\mathbf{\Theta}) }{{\cal Z}}
e^{-\left(\chi_{\rm SNIa}^2+\chi_{\rm
CMB}^2+\chi_{\rm BAO}^2+\chi_{\rm{GRO}}^{2} \right)/2}
\,,
\end{equation}
where the $\chi^2$ is a measure of the goodness of fit between the model theoretical predictions and the observed values of a certain physical quantity; ${\cal Z}~=~\int {\cal L}({\rm data}|\mathbf{\Theta}) \Pi(\mathbf{\Theta}) {\rm d}\mathbf{\Theta}$ is a normalization constant, irrelevant for parameter fitting.


\section{Results}

\subsection{Parameter estimation}

\subsubsection{Cosmological constant{\label{ccresults}}}

\begin{figure}[t!]
\begin{center}
\includegraphics[scale = 0.64]{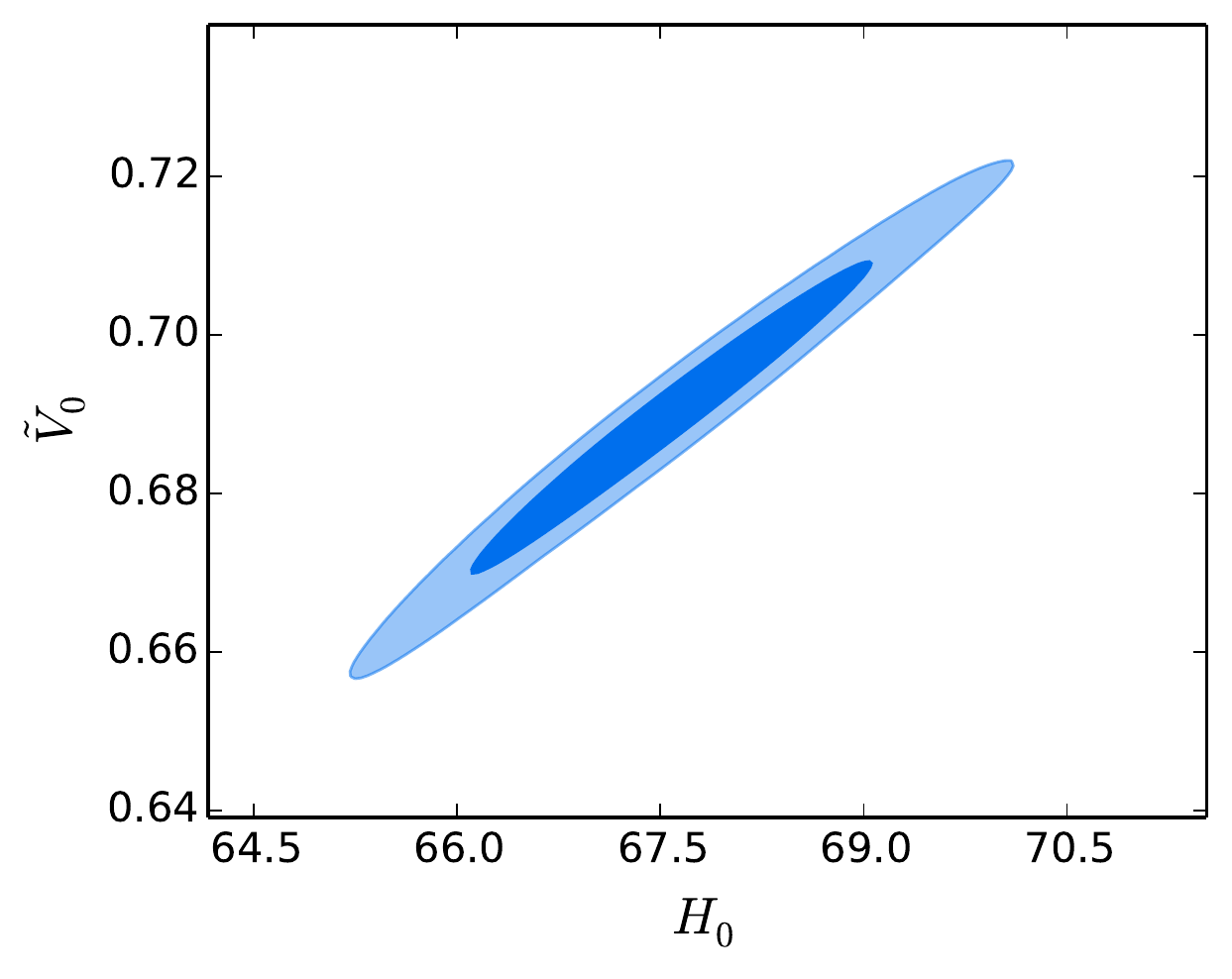}
\end{center}
\caption{\label{lamda} The $2$-d marginalized {\it Planck} constraints over the zeroth-order Taylor series potential model, representing an effective cosmological constant. We recover the results obtained by the {\it Planck} analysis for a flat Universe.}
\end{figure}

In this subsection, we show the marginalized $2$-d contour of the {\it Planck} constraints assuming a pure cosmological constant in our model. Therefore, we have set $V_i = 0$ for $i \neq 0$. Hence, $V_{0} \equiv \tilde{V}_{0}$ is just a derived parameter defined as the present-day energy density of the scalar field, with a value set to yield a flat cosmology. In Fig.~\ref{lamda}, we can see that our results match those from the {\it Planck} data release \cite{planckres}, as expected. 

\subsubsection{Linear potential \label{results_order1}}

\begin{figure*}[h!]
\begin{center}$
\begin{array}{c}
\includegraphics[scale = 0.32]{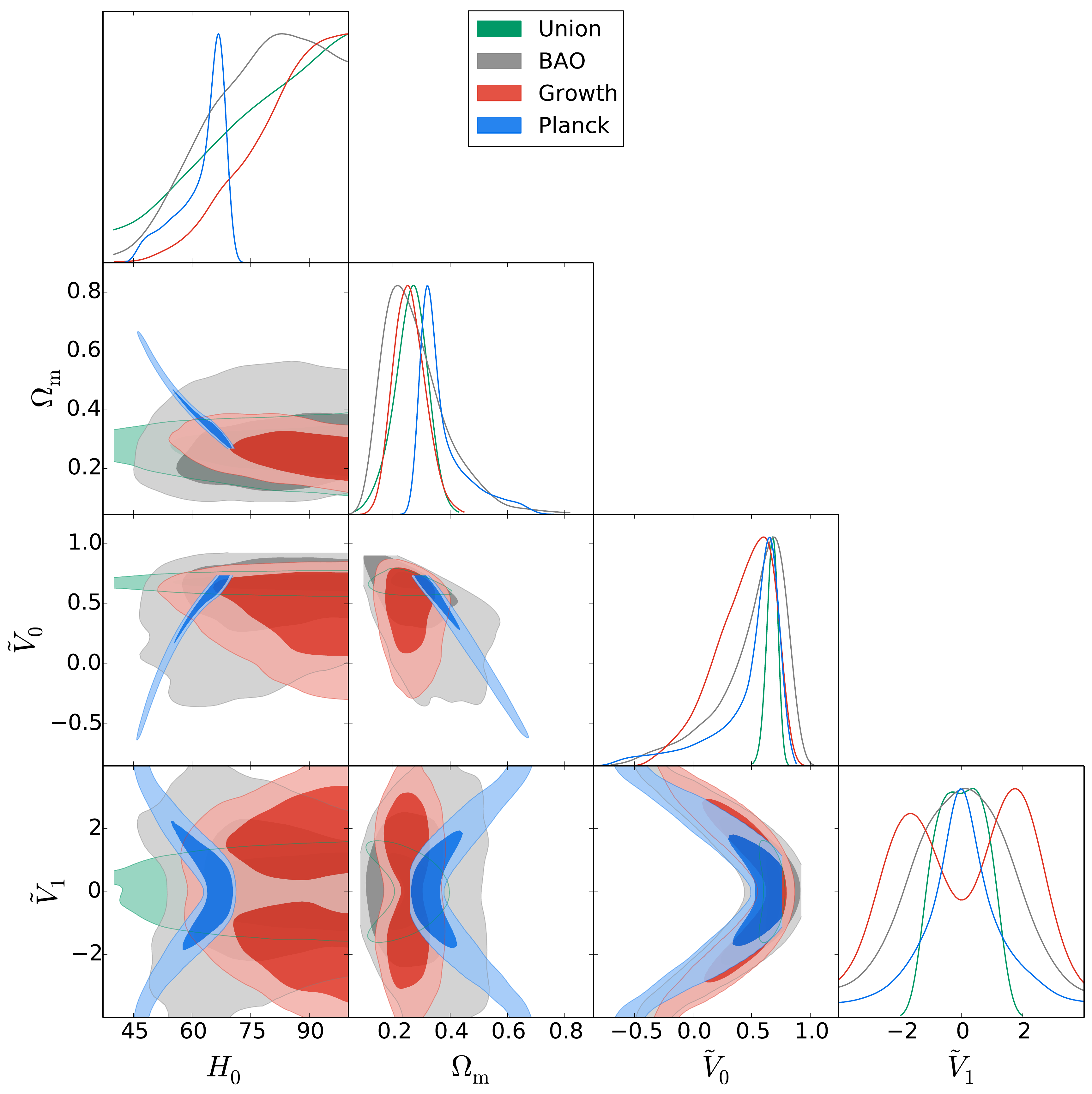} \\
\includegraphics[scale = 0.1935]{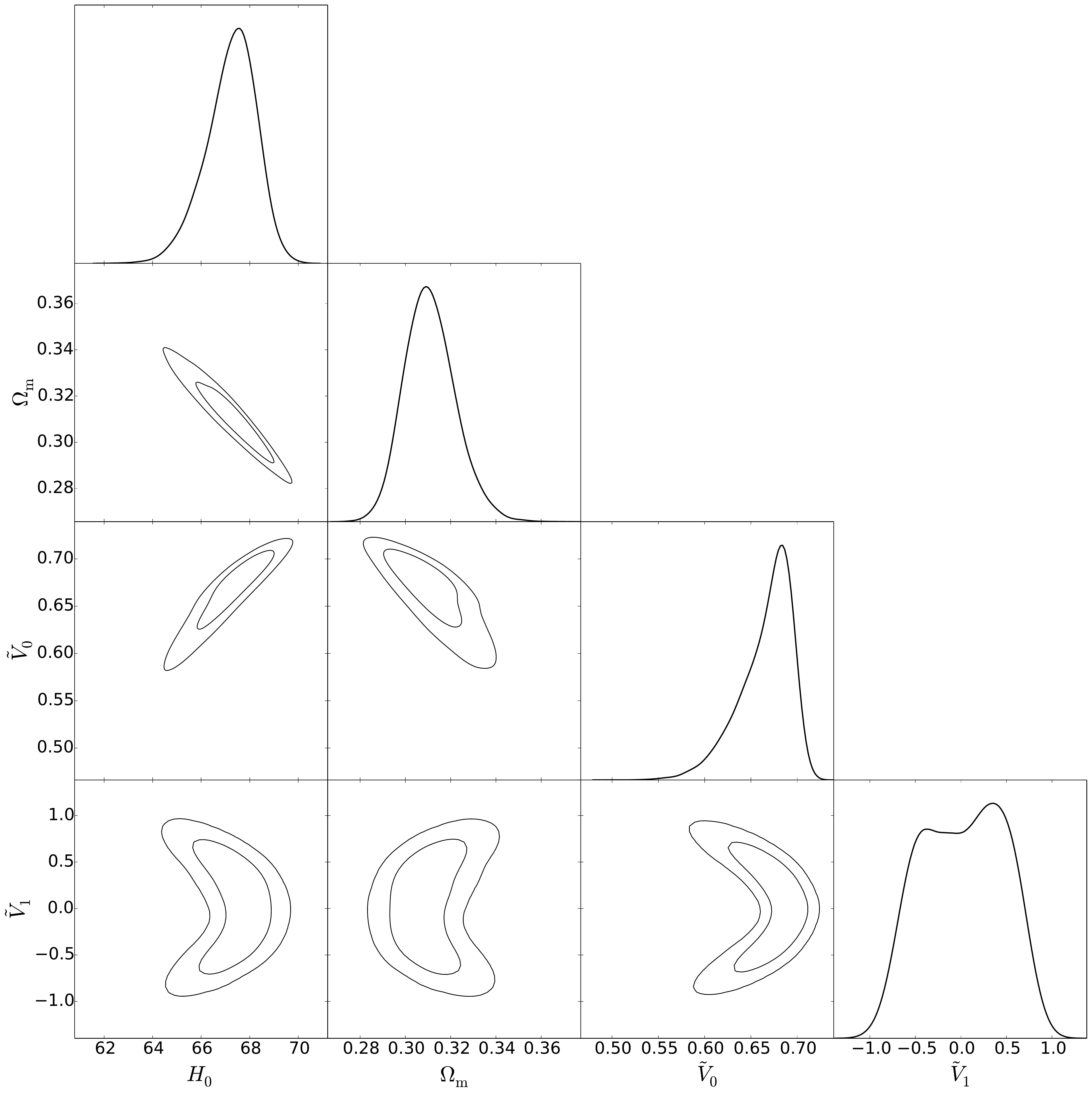}
\end{array}$
\end{center}
\caption{\label{order1_figure} The $2$-d contours and one-dimensional probability distributions for the first-order Taylor series potential model parameters. The top plot shows the constraints from the individual data sets (68\% and 95\% contours), while the bottom figure presents the constraints when all of the four data sets are combined (68\% and 95\% contours). The Planck contours include Planck, ACT and SPT data, as well as CMB lensing and $WMAP$ polarization.}
\end{figure*}

Here, we present the results for the linearized quintessence potential model, for which we have set $V_{2} = 0$. In Fig.~\ref{order1_figure} we show the marginalized $2$-d contours as well as the one-dimensional marginalized probability distributions for all the parameters considered in this model. We show the constraints from the individual data sets as well from their combination. 

In the top plot of Fig.~\ref{order1_figure} we have the constraints placed by the different data sets, individually, on the cosmological parameters, $H_{0}$ and $\Omega_{\rm{m}}$, and the dark energy parameters, $\tilde{V}_{0}$ and $\tilde{V}_{1} \equiv V_{1}$. The constraints on $\tilde{V}_{0}$ closely follow those obtained for $\Omega_{\rm{m}}$ since, as specified before, $V_{0}$ is actually a derived parameter, constrained from imposing a flat cosmology in our models. The larger uncertainty in $\tilde{V}_{0}$ is due to the inclusion of an additional parameter introducing new degeneracies in explaining the data. We note that the {\it Planck} data are clearly superior in constraining the traditional cosmological parameters (here $H_0$ and $\Omega_{\rm m}$), but that this superiority does not extend to the dark energy parameters.

\begin{figure}[t]
\begin{center}$
\begin{array}{c}
\includegraphics[scale = 0.385]{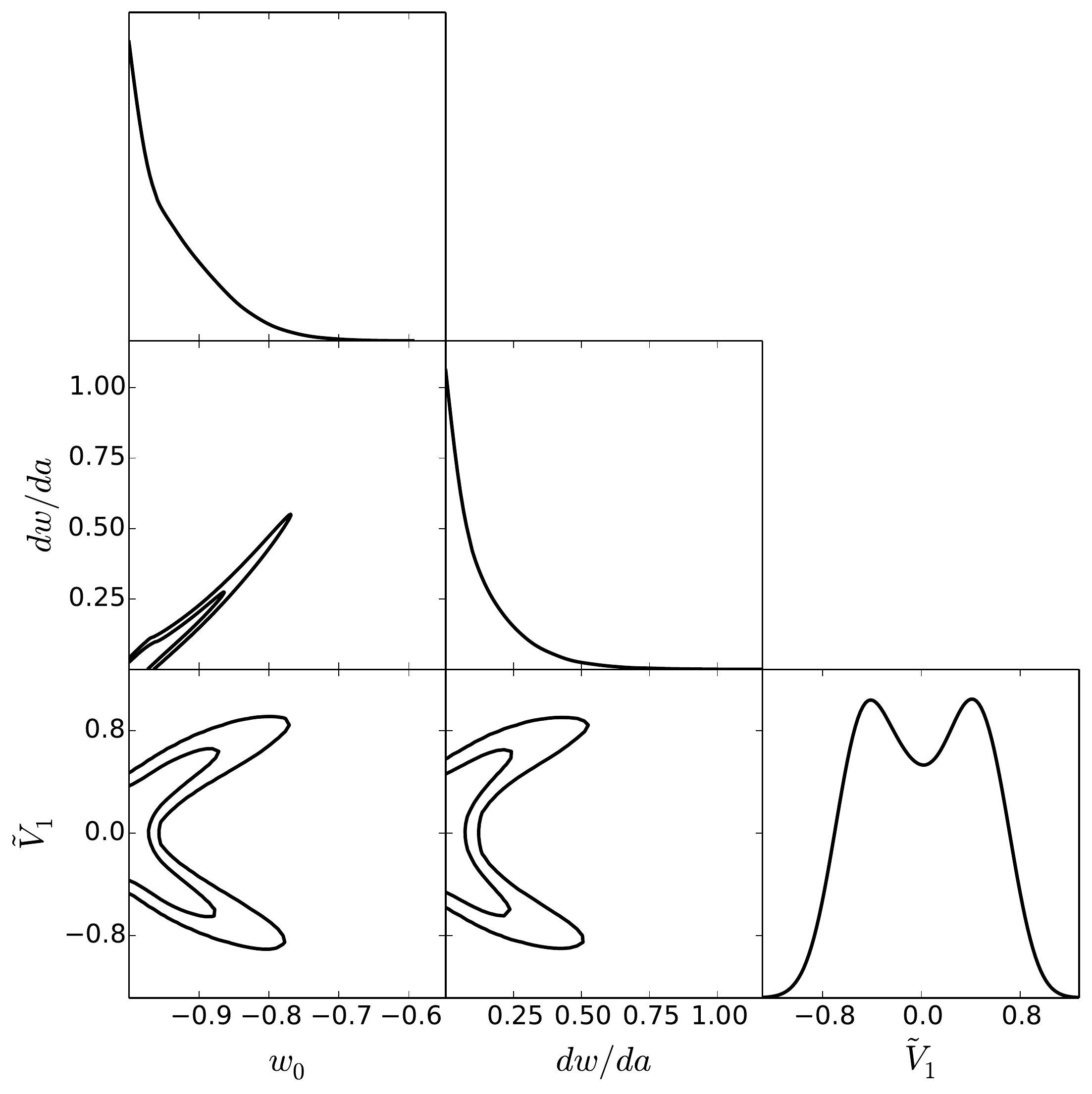}\\
\includegraphics[scale = 0.32]{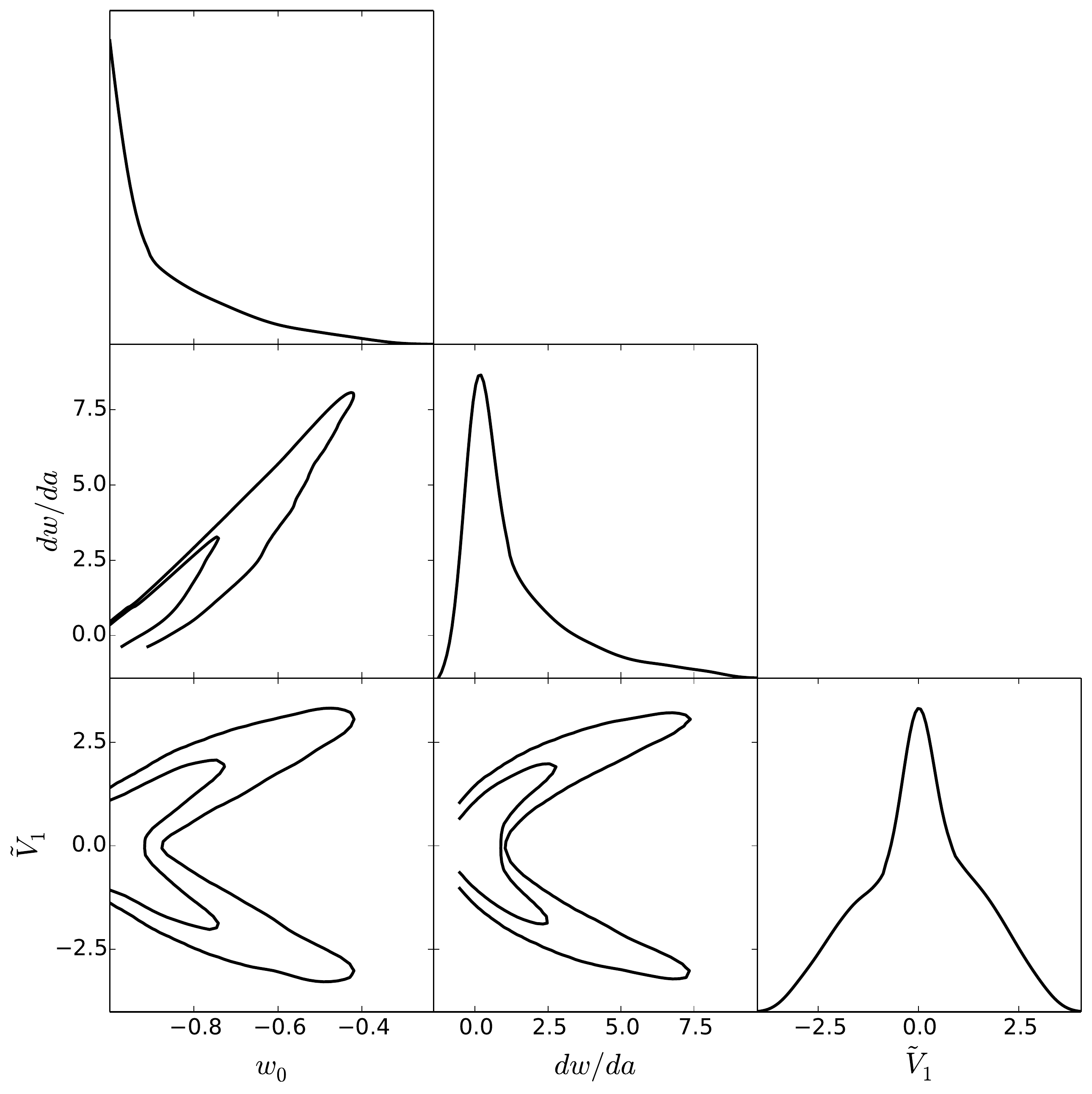}
\end{array}$
\end{center}
\caption{\label{w0waorder1} The $2$-d contours in the $w_0$--$w_a$ plane and the respective marginalized posterior probabilities distributions for the first (upper plot) and second-order (bottom plot) Taylor series potential. $dw/da \equiv -w_a$ was computed at $z = 0$. Since the x-axis limits are so different between the different orders of the potential expansion, we do not join both plots.}
\end{figure}

The supernova data are the most effective in constraining the DE parameters, with the smallest limits on the posterior probability distribution of $\tilde{V}_{1}$. The other datasets give comparable performance to each other. 

\begin{table*}[ht]
 \caption{68\% confidence limits of the first and second-order potential model parameters from the different data sets considered and their combination. We present the $95 \%$ upper limit on the absolute value of $\tilde{V}_{1}$, given its reflection symmetry. We don't show confidence ranges for $\tilde{V}_{2}$ as the models at the prior edges are not considerably disfavored compared to the best-fit.\label{order1_table}}
\begin{center}
\begin{tabular}{|c|c|c|c|c|c|c|}
\hline
   & & $H_{0}$ & $\Omega_{\rm{m}}$ & $\tilde{V}_{0}$ & $|\tilde{V}_{1}|$ & $\tilde{V}_{2}$ (best-fit) \\ \hline
   & Planck  & $61 \pm 6$ & $0.39 \pm 0.09$ & $0.4 \pm 0.3$  & $|\tilde{V}_{1}| < 3.3$ &\\ 
  & Union  & $78 \pm 15$  & $0.26 \pm 0.06$ & $0.67 \pm 0.05$ & $|\tilde{V}_{1}| < 1.3$ & \\
  Linear & BAO & $79 \pm 13$  & $0.29 \pm 0.12$ & $0.5 \pm 0.3$ & $|\tilde{V}_{1}| < 3.2$ & \\
   & Growth & $85 \pm 11$  & $0.26 \pm 0.06$  & $0.41 \pm 0.25$ & $|\tilde{V}_{1}| < 3.3$ & \\  
   & Combined & $67.2 \pm 1.0$ & $0.311 \pm 0.011$ & $0.66 \pm 0.03$ & $|\tilde{V}_{1}| < 0.77$ & \\ \hline
  & Planck & $66 \pm 3$  & $0.33 \pm 0.04$ & $0.60 \pm 0.10$ & $|\tilde{V}_{1}| < 2.8$ &  \\
  & Union & $79 \pm 15$ & $0.27 \pm 0.05$ & $0.68 \pm 0.06$ & $|\tilde{V}_{1}| < 2.5$ &   \\
   Quadratic & BAO & $82 \pm 13$ & $0.25 \pm 0.10$ & $0.68 \pm 0.13$ & $|\tilde{V}_{1}| < 2.7$ & \\
  & Growth & $80 \pm 12$ & $0.27 \pm 0.06$ & $0.64 \pm 0.11$ & $|\tilde{V}_{1}| < 2.8$ & \\
  & Combined & $67.2 \pm 1.1$ & $0.311 \pm 0.012$ & $0.64 \pm 0.06$ & $|\tilde{V}_{1}| < 2.7$ &  $-7$ \\ \hline 
  
\end{tabular}
\end{center}
\end{table*}

Combining the four data sets, as seen on the lower plot of Fig.~\ref{order1_figure}, the confidence limits improve considerably. The confidence limits resulting from the four data sets combined can be seen in Table \ref{order1_table}. Nevertheless, the improvement is modest compared to previous works of Refs.~\cite{2005PhRvD..72h3511S, 2007PhRvD..75b3502S}, with most of the constraining power coming from the Supernovae alone and the contributions from the other data sets are not particularly enhancing. As expected the posterior probability distribution of $\tilde{V}_{1}$ is symmetric around zero, the perfect cosmological constant case, which is well within the 68\% confidence limits.

It is interesting to see how the individual data sets behave and how they combine, particularly in the $\Omega_{\rm{m}} - \tilde{V}_{1}$ plane. We observe that the constraints from Supernovae and {\it Planck} exhibit opposite trends in how the preferred $|V_1|$ changes with $\Omega_{\rm m}$. In the Supernovae case, a tilted potential compensates a smaller matter energy density in the luminosity distance because a higher tilt corresponds, effectively, to a higher dark energy equation of state (i.e.\ higher than $-1$). Hence, the dark energy contribution at higher redshift increases, compensating for a smaller $\Omega_{\rm{m}}$, leading to the left-sided boomerang shape we observe.

As for the {\it Planck} constraints, the distance-related observables considered are the acoustic scale, $l_{A}$, and the shift parameter $R$ \cite{mukherjee}. $l_{A}$ corresponds to the ratio between the angular-diameter distance, $D_{A}$, and the comoving sound horizon, $r_{\rm{s}}$, at photon decoupling. Increasing $\Omega_{\rm{m}}$ decreases both quantities, and a tilted potential also decreases $D_{A}$ due to a higher dark energy effective equation of state, compensating for higher matter energy densities. Lastly, $R$ is the product between $\Omega_{\rm{m}}$ and $D_{A}$. Increasing the matter density is, therefore, compensated by a higher $V_1$. All these effects end up contributing for the right-shaped boomerang contour we observe.

A further aspect worth discussing is the clear bimodality in the $H_{0}$--$\tilde{V}_{1}$ contours produced by the growth data (also present in the $\Omega_{\rm{m}}$--$\tilde{V}_{1}$ contours) in Fig.~\ref{order1_figure}. 
As an increase in $H_{0}$ increases $\sigma_8$ by altering the shape of the power spectrum, a higher density of dark energy is required in order to remain consistent with the growth measurements of $f \sigma_8$, through lowering $f$.

Finally, in Fig.~\ref{w0waorder1} we see the constraints obtained in the $w_0$--$w_a$ plane, where $dw/da \equiv -w_{a}$ was computed at $z = 0$. We note how our results restrict themselves to a thawing regime, with the scalar field becoming free at later times to roll down the potential and increasing the effective equation of state when $V_{1} \neq 0$. Hence our constraints are restricted to the $w_{0} \geq -1$ and $w_a \leq 0$, with the data preferring the $\Lambda$CDM case, as expected.

\subsubsection{Quadratic potential \label{results_order2}}

\begin{figure*}[t!]
\begin{center}$
\begin{array}{c}
\includegraphics[scale = 0.26]{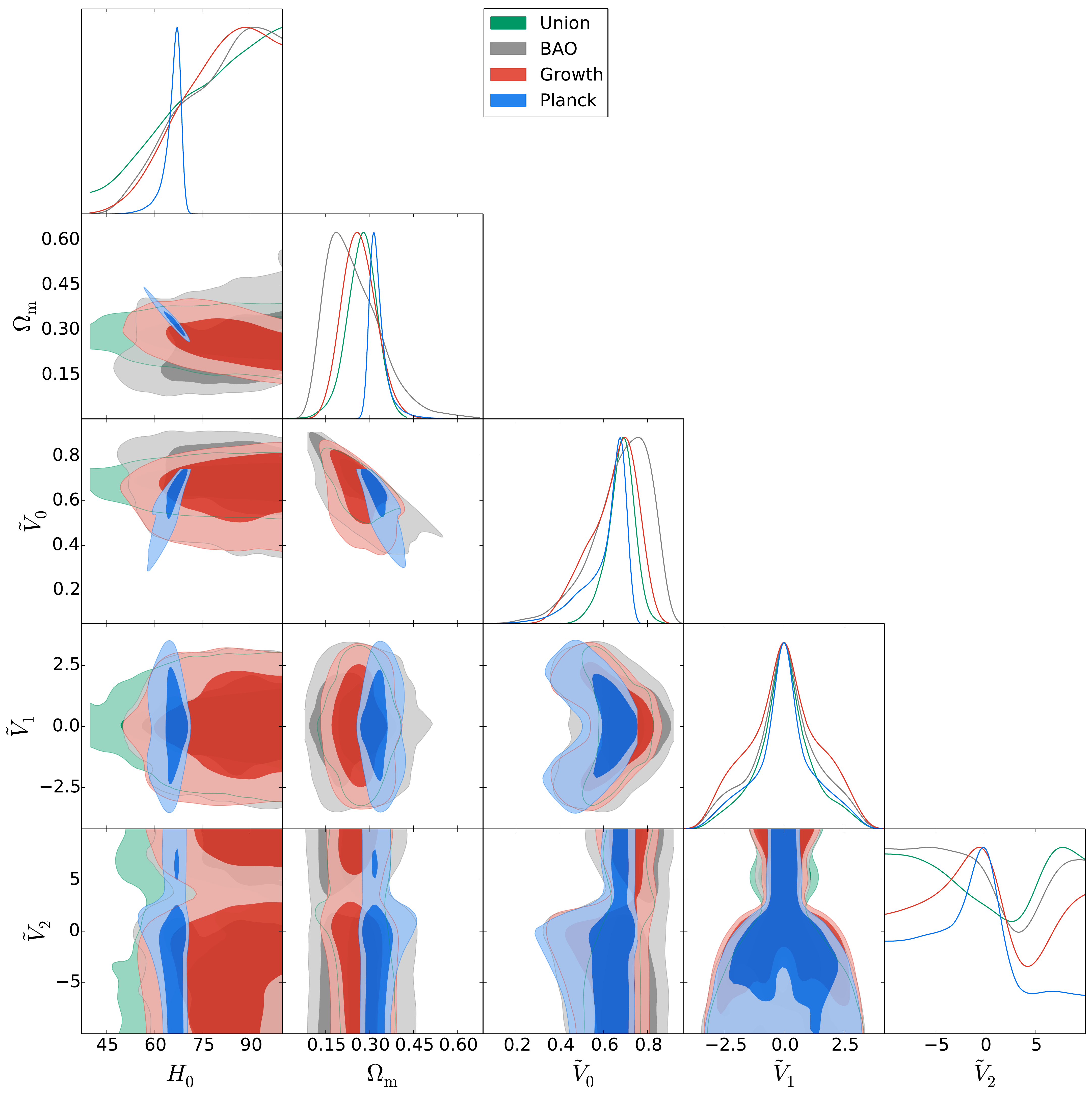} \\
\includegraphics[scale = 0.26]{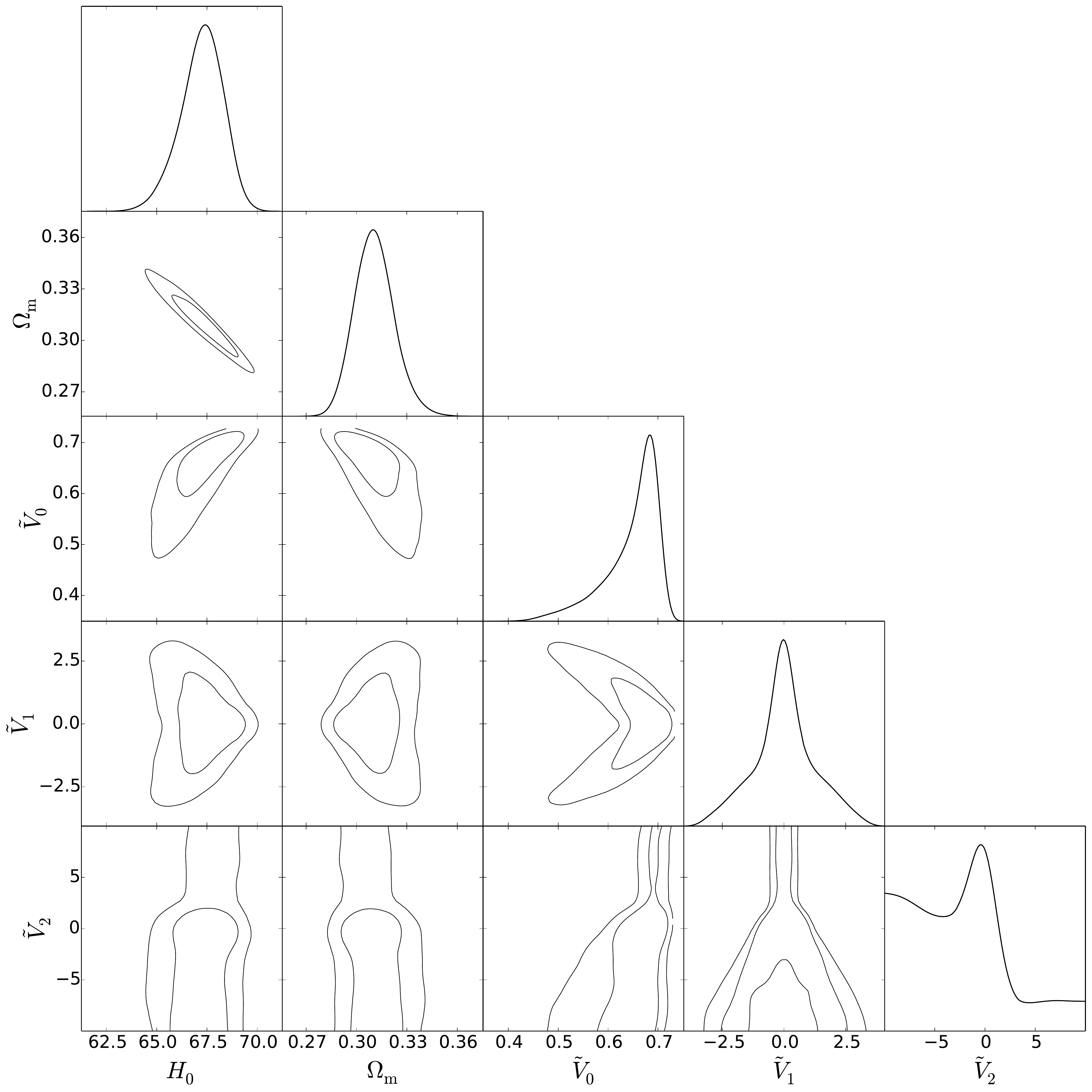}
\end{array}$
\end{center}
\caption{\label{order2_figure} As Fig.~\ref{order1_figure}, for the second-order Taylor series potential model parameters.}
\end{figure*}

In this subsection, we show the results for the quadratic quintessence potential model, where we vary both $V_{1}$ and $V_{2}$. In Fig.~\ref{order2_figure} we have the marginalized $2$-d contours as well as the one-dimensional marginalized probability distributions of all the parameters considered. We show the constraints from the individual data sets as well from their combination. The confidence limits of the parameters that define this model can be seen in Table \ref{order1_table}.

Inevitably the constraints over our parameters get worse by considering a higher order of the potential. As in the previous case, the Supernovae data performs better (although now only marginally better) in constraining the dark energy parameters over the other three available data sets. The confidence limits in $\tilde{V}_{0}$ and $\tilde{V}_{1}$ almost doubled compared to the first-order potential counterparts. 
The limits on the $1$-d posterior probability distribution of $\tilde{V}_{2}$ extend to the prior we have defined for this parameter on all the data sets considered, in a clear indication of the difficulty of the data in constraining it. Our understanding is that the limits would always be cut by the prior we would define, had it been smaller or larger. For instance, the positive, upper limit would extend indefinitely, as the data will always accept models with a large $\tilde{V}_{2}$ as long as $\tilde{V}_{1}$ remains zero. This is because, in our search for the initial conditions, that means the field will remain immobile at the minimum ($\phi = 0$) of the convex potential, producing, effectively, a $\Lambda$CDM-like model, which is in strong agreement with the data. Hence, one obtains the narrow region extending to high values of $\tilde{V}_{2}$ in the $2$-dimensional $\tilde{V}_{1} - \tilde{V}_{2}$ contours.

As for the bifurcated shape in the $\tilde{V}_{1} - \tilde{V}_{2}$ contours when considering the negative region of $\tilde{V}_{2}$, particularly for smaller, negative values of it, is due to the fact that it is increasingly difficult to obtain a zero-valued $\tilde{V}_{1}$ in these conditions. The concave nature of the potential makes it hard for the field to remain still at $\phi = 0$, as the smallest tilt will make the field roll considerably. Hence, the displacement of the field will always produce a non-zero value of $\tilde{V}_{1}$, creating the shape we can see in Fig.~\ref{order2_figure}. This could extend indefinitely according to the negative prior limit on $\tilde{V}_{2}$, with the increasing displacement of the field producing a widening of the legged shape figure.

In Fig.~\ref{w0waorder1} we have the constraining contours on the \mbox{$w_0$ -- $w_a$} plane for the second-order Taylor expanded potential. We observe something very similar to the linear potential results, with our results constrained to a thawing regime. We do have now a very small region of $w_{a} > 0$, which is however not indicative of a traditional freezing regime. This actually is a result of the field rolling up the potential close to $z = 0$ after passing its minimum and being slowed down by the slope of the potential. This type of behavior is observed, for instance, in the models considered in Ref.~\cite{thawing1}. We also note the very different $w_0$--$w_a$ ranges between the two models, as the quadratic potential is able to cover much more of this plane with observationally-permitted models.

Lastly, we also observe an odd feature in the $\tilde{V}_{2}$--$H_0$ contour plots of Fig.~\ref{order2_figure}. For $\tilde{V}_{2} > 0$, we note that the allowed values of $\tilde{V}_{1}$ get significantly smaller. Hence, this shrinks $\tilde{V}_{0}$ and $\Omega_{\rm{m}}$ towards larger and smaller values respectively, with  $\Omega_{\rm{m}}$ affecting the allowed values of $H_{0}$ (towards larger values) due to the known degeneracy between these two parameters. The same reasoning can be applied for $V_{2} \lesssim 0$: $\tilde{V}_{1}$ can assume larger values, widening $\tilde{V}_{0}$ and $\Omega_{\rm{m}}$ towards smaller and larger values respectively, making $H_{0}$ tend to smaller values and producing a bump in the $\tilde{V}_{2}$--$H_0$ contour plot.

\section{Conclusions}

In this work, we expanded the quintessence scalar field potential in a Taylor series around the field's present-day value, which we set to zero by a rescaling of the potential expansion parameters. We also limit ourselves to a thawing scenario by setting the field's initial velocity to zero. We then used CosmoMC to constrain these parameters with the latest available data.

The main addition in this study was the inclusion of growth of structure data, which constrains the models against observations of redshift distortions of objects and measurements of the normalized growth rate, $f \sigma_{8}(z)$. We conclude that the growth data performance in constraining the dark energy parameters we have considered is quite competitive compared to the other three data sets we have used, with the supernova data providing the tightest constraints on the potential parameters.

We have also shown explicitly how the different data sets constrain our model parameters. It is particularly interesting how the Supernovae and the {\it Planck} contours behave in opposite manner in the $\Omega_{\rm{m}}-V_{1}$ plane. This could be promising for future surveys, proving to be the ideal combination for constraining dynamical dark energy models due to the observables involved.

The improvement in constraints we observe is modest relative to our previous works \cite{2005PhRvD..72h3511S, 2007PhRvD..75b3502S}, where the data used was the {\it WMAP} $3$-year distance information data \cite{wmap3}, the $\rm{SNLS}$ type Ia Supernova dataset \cite{snlsref}, and the SDSS and 2dF galaxy redshift survey baryon acoustic peak information \cite{sdssmartin,2dfmartin}. It is true that the model we consider is different by construction, since we restrict ourselves to thawing quintessence models, but nonetheless the constraints on the parameters of the expanded potential did not improve significantly in the many years since those works despite the accumulated data of the past decade, both of existing types and new growth of structure data. However it is notable that those early works had a fairly narrow prior on $V_1$ which contributed to the quoted strength of the constraint, whereas our wider prior shows this parameter to now be well constrained by data in the linear potential case.

A comparable example is the marginal improvement of constraints on the DE equation of state, $w$, or on the $2$-dimensional contours on the $w_0$--$w_a$ plane (we suggest comparing the results from the {\it WMAP} $9$-year data with the {\it Planck} results \cite{wmap9,planckres}). It has been shown that the $w_0$--$w_a$ parameterization is a good representation of thawing quintessence models \cite{linder1}. Therefore, given the expectation that next-generation experiments will be unable to distinguish thawing quintessence with $w_{0} < -0.9$ from $\Lambda$CDM at the $95\%$ confidence level \cite{linder2}, we predict that improvement on the constraints we show in this work to be challenging also.


Despite this, we conclude that for the first-order Taylor series expansion of the potential, $\Lambda$CDM is well within the preference of the data, with the confidence limits on the parameters we consider being $\tilde{V}_{0} = 0.66 \pm 0.03$ and $|\tilde{V}_{1}| < 0.77 \hspace{1 mm} (95 \%)$. These constraints get significantly worse when we consider an extra order on the expansion of the potential. The second-order term, $\tilde{V}_{2}$,  is unconstrained by the data, with the posterior limits on this parameter extending to the prior range we define for it, possibly indefinitely. 


\begin{acknowledgments}
N.A.L. was supported by Funda\c{c}\~{a}o para a Ci\^{e}ncia e Tecnologia (FCT) through grant SFRH/BD/85164/2012. A.R.L.\ was supported by the Science and Technology Facilities Council [grant numbers ST/K006606/1 and ST/L000644/1]. M.S.\ was supported by the Templeton Foundation. D.R.P. was supported by an Australian Research Council Future Fellowship [grant number FT130101086].
This work was undertaken on the COSMOS Shared Memory system at DAMTP, University of Cambridge operated on behalf of the STFC DiRAC HPC Facility. This equipment is funded by BIS National E-infrastructure capital grant ST/J005673/1 and STFC grants ST/H008586/1, ST/K00333X/1.
N.A.L. would also like to thank Vanessa Smer-Barreto for many helpful discussions about CosmoMC.
\end{acknowledgments}


\bibliography{recquint3.bib}

\end{document}